\documentclass[sigconf]{acmart}
\AtBeginDocument{%
  \providecommand\BibTeX{{%
    Bib\TeX}}}

\usepackage{graphicx}
\usepackage{subfiles}

\usepackage{todonotes}
\usepackage{booktabs}
\PassOptionsToPackage{hyphens}{url}\usepackage{hyperref}

\usepackage{dsfont}
\usepackage{amsmath}

\usepackage{caption}
\usepackage{subcaption}

\usepackage{xcolor}
\usepackage{xspace}

\usepackage{pifont}% http://ctan.org/pkg/pifont

\newcommand{\argue}{\textsc{ARGUE}\xspace}
\newcommand{\aargue}{\textsc{Auto-ARGUE}\xspace}
\newcommand{\argueviz}{\textsc{ARGUE-viz}\xspace}
\newcommand{\autonugget}{\textsc{AutoNuggetizer}\xspace}

\AtBeginDocument{%
  \providecommand\BibTeX{{%
    \normalfont B\kern-0.5em{\scshape i\kern-0.25em b}\kern-0.8em\TeX}}}

\copyrightyear{2026}
\acmYear{2026}
\setcopyright{cc}
\setcctype{by}
\acmConference[SIGIR '26]{Proceedings of the 49th International ACM SIGIR Conference on Research and Development in Information Retrieval}{July 20--24, 2026}{Melbourne, VIC, Australia}
\acmBooktitle{Proceedings of the 49th International ACM SIGIR Conference on Research and Development in Information Retrieval (SIGIR '26), July 20--24, 2026, Melbourne, VIC, Australia}
\acmDOI{10.1145/3805712.3808384}
\acmISBN{979-8-4007-2599-9/2026/07}

\begin{document}

\title[\aargue: LLM-Based Report Generation Evaluation]{\aargue: \\ LLM-Based Report Generation Evaluation}

\settopmatter{authorsperrow=5}

\author{William Walden}
\email{wwalden1@jh.edu}
\affiliation{
\institution{Johns Hopkins University}
\city{Baltimore}
\state{MD}
\country{USA}
}
\orcid{1234-5678-9012}

\author{Marc Mason}
\email{mmason@solerity.edu}
\affiliation{
\institution{Solerity}
\city{Herndon}
\state{VA}
\country{USA}
}

\author{Orion Weller}
\email{oweller2@jh.edu}
\affiliation{
\institution{Johns Hopkins University}
\city{Baltimore}
\state{MD}
\country{USA}
}

\author{Laura Dietz}
\email{dietz@unh.edu}
\affiliation{
\institution{University of New Hampshire}
\city{Durham}
\state{NH}
\country{USA}
}

\author{John M.\ Conroy}
\email{conroy@super.org}
\affiliation{
\institution{IDA Center for Computing Services}
\city{Bowie}
\state{MD}
\country{USA}
}

\author{Neil Molino}
\email{npmolin@super.org}
\affiliation{
\institution{IDA Center for Computing Services}
\city{Bowie}
\state{MD}
\country{USA}
}

\author{Hannah Recknor}
\email{hreckno1@jh.edu}
\affiliation{
\institution{Johns Hopkins University}
\city{Baltimore}
\state{MD}
\country{USA}
}

\author{Bryan Li}
\email{bryanli.ca@gmail.com}
\affiliation{
\institution{Google}
\city{New York}
\state{NY}
\country{USA}
}

\author{Gabrielle Liu}
\email{kaili.liu@yale.edu}
\affiliation{
\institution{Yale University}
\city{New Haven}
\state{CT}
\country{USA}
}

\author{Yu Hou}
\email{houyu@umd.edu}
\affiliation{
\institution{University of Maryland}
\city{College Park}
\state{MD}
\country{USA}
}

\author{Dawn Lawrie}
\email{lawrie@jhu.edu}
\affiliation{
\institution{Johns Hopkins University}
\city{Baltimore}
\state{MD}
\country{USA}
}

\author{James Mayfield}
\email{mayfield@jhu.edu}
\affiliation{
\institution{Johns Hopkins University}
\city{Baltimore}
\state{MD}
\country{USA}
}

\author{Eugene Yang}
\email{eugene.yang@jhu.edu}
\affiliation{
\institution{Johns Hopkins University}
\city{Baltimore}
\state{MD}
\country{USA}
}

% \affiliation{%

\renewcommand{\shortauthors}{Walden et al.}

\begin{abstract}
  Generation of citation-backed reports is a primary use case for retrieval-augmented generation (RAG) systems. While open-source evaluation tools exist for various RAG tasks, tools designed for report generation are lacking. Accordingly, we introduce \aargue, a robust LLM-based implementation of the recently proposed \argue framework for report generation evaluation. We present analysis of \aargue on the report generation pilot task from the TREC 2024 NeuCLIR track and on two tasks from the TREC 2024 RAG track, showing good system-level correlations with human judgments. Additionally, we release \argueviz, a web app for visualization and fine-grained analysis of \aargue judgments and scores.\footnote{\aargue: \url{https://github.com/hltcoe/auto-argue}; \url{https://youtu.be/M9ODL6j05B0}\\\phantom{$^1$}\argueviz: \url{https://github.com/hltcoe/argue-viz}; \url{https://youtu.be/GAZFljTLW1c}}
\end{abstract}





\maketitle

\section{Introduction}
\label{sec:intro}
As applications of retrieval-augmented generation (RAG) systems have proliferated in recent years, interest in methods for automatic evaluation of RAG outputs has grown in tandem. Numerous works have positioned themselves as \emph{general} solutions for RAG evaluation \cite{es2024ragas,gao2023enabling,pradeep2025great,saad-falcon-etal-2024-ares}, and while some evaluation desiderata are shared across tasks, many important, \emph{task-dependent} considerations persist.

Our work focuses on \emph{report generation}, a RAG task that aims to produce a long-form, citation-attributed response to a complex query. At least two key features distinguish report generation from related RAG tasks, such as long-form QA. First, report generation strongly foregrounds the identity of the user (or \emph{requester}): the same query should in principle yield different reports for requesters with different backgrounds (e.g., levels of education or domain expertise). Second, the ideal report represents a \emph{summary} over the entire corpus of the most user-critical information, thus stressing full \emph{coverage} of the collection where traditional QA emphasizes mere \emph{sufficiency} of the response.

\begin{figure*}
    \centering
\includegraphics[width=0.8\textwidth]{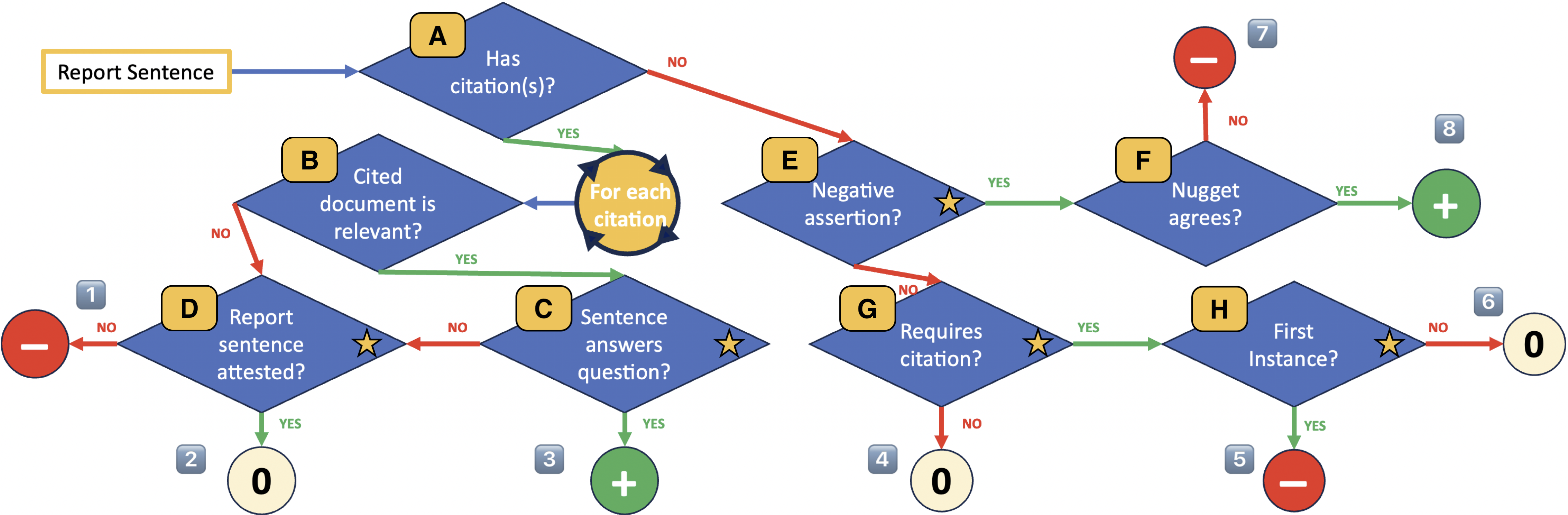}
    \caption{The ARGUE framework from Mayfield et al.\ \cite{mayfield2024evaluation}, adapted with permission from the authors.}
    \label{fig:argue}
    % \vspace{-7mm}
\end{figure*}

Many evaluation methods for RAG cover certain facets of report generation in isolation---e.g., recall of key pieces of information or \emph{nuggets} \cite{lajewska2025ginger,pradeep2025great,dietz2026incorporating,dietz2026insider}. In contrast, the \argue framework \cite{mayfield2024evaluation} from \citeauthor{mayfield2024evaluation} represents one of the most integrated proposals for full report generation evaluation---emphasizing not only information coverage but also proper attribution of claims, as well as correct assertions of ignorance when certain aspects of the query cannot be addressed. A key limitation of their work, however, is the lack of any open-source \emph{automatic} implementation of \argue.

%inhibiting its application to real RG tasks.

Our work addresses this limitation by introducing \textbf{\aargue}, the first public implementation of \argue. \aargue uses the LLM-as-a-judge paradigm to automate the sequence of binary, sentence-level judgments that the framework requires (\autoref{fig:argue}) and from which it computes report-level scores for key dimensions of report quality. \aargue is highly configurable, enabling use of different collections, LLM judges, model providers, and nugget sets, and---critically---facilitates reproducible evaluations.

\argue is complex, however, and typically produces many judgments for each report. Although top-line metrics give some sense of a report's quality, they do not permit detailed error analysis.

Accordingly, alongside \aargue, we release \textbf{\argueviz}, a visualization tool that presents fine-grained information about nugget coverage and citation support as judged by \aargue in a simple web interface. \argueviz thus enables localization of system errors and faster diagnosis of system limitations.

Finally, we present evidence for the effectiveness of \aargue through two sets of experiments. First, we demonstrate strong system-level correlations between \aargue scores and assessor-annotated \argue scores on runs from the TREC 2024 NeuCLIR report generation pilot task. And second, we show that \aargue correlates well with human judgments of nugget support derived from the \autonugget framework \cite{pradeep2025great} on the TREC 2024 RAG report generation task.

% We summarize our contributions as follows:
% \begin{enumerate}
%     \item \textbf{\aargue}: An automatic, LLM-based Python implementation of the \argue framework---the only publicly available \argue implementation.
%     \item \textbf{\argueviz}: A web app for fine-grained visualization and analysis of \aargue outputs.
%     \item \textbf{Case Study}: Meta-evaluation results with \aargue on the TREC 2024 NeuCLIR report generation pilot task \cite{lawrie2025overview}.
% \end{enumerate}
% \aargue is configurable, easy-to-use, and extensible to other RG datasets, adopting the new TREC format for RAG outputs\footnote{Format schema and validator: \url{https://github.com/hltcoe/rag-run-validator}}. We release \aargue\footnote{\url{https://github.com/hltcoe/auto-argue}} and \argueviz\footnote{\url{https://github.com/hltcoe/argue-viz}} to facilitate further work on automatic RG evaluation.

\section{\argue}
\label{sec:argue}
% \vspace{-8mm}
\autoref{fig:argue} depicts the \argue framework, which evaluates reports via a tree of binary judgments (blue diamonds) about each sentence's content and citations. Depending on the path(s) traversed, a report may incur penalties (red circles), rewards (green), or neither (beige).

\paragraph{Inputs.} \argue takes as input: (1) a \emph{report}; (2) the \emph{report request}, comprising a \emph{problem statement} that describes the information need and a \emph{user story} that describes the requester; (3) the document collection used to generate the report; and (4) a collection of \emph{nuggets} (question-answer pairs), optionally with a mapping between nugget answers and documents that attest them.

\paragraph{Content: Nuggets.} Following prior work \cite{alaofi2024generative,lin2005automatically,pradeep2025great,rajput2011nugget,voorhees2003overview}, \argue evaluates reports' coverage of relevant information via sets of \emph{nuggets}---questions that an ideal report should address, paired with answers linked to documents that attest them.\footnote{We will use the terms \emph{nugget question} and \emph{nugget answer} to refer the question and answer components (respectively) of a given nugget.} Each report sentence is assessed to determine which nugget question(s) it answers correctly and reports are rewarded for each such nugget.

\paragraph{Citations.} Citations are assumed to support only the sentence they follow.\footnote{While it is common for citations to be somewhat lazily clustered at the end of several related sentences---or even a paragraph---this assumption encourages more rigorous citation practices and makes automated evaluation more straightforward.} Sentences may bear 0+ citations. Sentences supported by their citations are rewarded; those that have non-supporting citations, or that require citations but lack them, are penalized.

\paragraph{Edge Cases.} \argue also handles two edge cases not covered by other frameworks. The first relates to repeated information (\autoref{fig:argue}, Judgment H): if a sentence lacks citations, it is penalized only if it presents \emph{new} information (outcome 5); reiterating an earlier claim without citation carries no penalty (outcome 6). The second relates to assertions of ignorance, dubbed \emph{negative assertions} (Judgment E): sentences claiming that \emph{no} answer can be found for some aspect of the report request are validated against a corresponding nugget (Judgment F). These sentences are rewarded if and only if the nugget has an empty answer set and are penalized otherwise.

\paragraph{Metrics.} \argue does not prescribe specific metrics; it merely produces \emph{judgments} from which various metrics may be computed. Nonetheless, Mayfield et al.\ recommend two metrics---\emph{sentence precision} and \emph{nugget recall}---which we discuss in \S\ref{sec:auto-argue}.

% We refer readers to \cite{mayfield2024evaluation} for further details on \argue.

\section{\aargue}
\label{sec:auto-argue}
The \argue framework leaves a great deal to implementation, including the judge (whether human or machine), the magnitudes of rewards and penalties, the source of the nuggets and relevance labels, and the metrics to be computed. Below, we detail the choices we made in implementing \aargue.

\paragraph{Judge.} An LLM judge is queried using few-shot prompts to obtain binary (YES/NO) answers to all non-trivial judgments (starred in \autoref{fig:argue}) for each report sentence. Answers to the remaining judgments (e.g., whether a sentence has citations) are evaluated deterministically via lookup.

\paragraph{Relevance Judgments.} \argue makes a relevance assessment of each cited document in a report (Judgment B in \autoref{fig:argue}). In \aargue, a document is deemed relevant if and only if it attests \emph{at least one} nugget answer, determined either via lookup in a document-nugget mapping (see below) or via a call to the judge.\footnote{In our experiments (\S\ref{sec:experiments}), we perform this check via lookup, although \aargue supports both methods.}

\paragraph{Nuggets.} Nuggets may have multiple answers (each attested by 1+ documents) and may come in one of two varieties: \textsc{AND} nuggets, for which \emph{all} answers must be given, and \textsc{OR} nuggets, for which only \emph{one} answer is required. Nuggets may also have importance labels---\texttt{vital} or \texttt{okay}---which carry different weights (see \emph{Metrics}).

\paragraph{Nugget Attestation} If a cited document is deemed relevant, \argue then requires assessing which nuggets are attested by the citing sentence (\autoref{fig:argue}, Judgment C). Given the loop over citations for each sentence, an implementation based na{\"i}vely on \autoref{fig:argue} would make one call to the judge per cited document ($N_{cite}$), per nugget answer ($N_{ans}$)---yielding $N_{cite} \times N_{ans}$ total calls.

We make two optimizations to substantially reduce this number. First, we rely on a mapping from documents in the collection to the nugget answers they attest. This mapping is generally obtained via the nugget annotation process, which involves creating nuggets directly from (clusters of) documents. Thus, for each sentence, the mapping enables us to run the attestation check only for the subset of nugget answers attested by \emph{at least one} of the documents cited by that sentence.\footnote{While this optimization is the default behavior, \aargue still supports running the attestation check for \emph{all} nuggets if desired.} Second, because the attestation check is a function only of a sentence and a nugget answer, we need not repeat the check for each citation ($N_{cite}$) and instead perform it only once per answer that is attested by at least one of the cited documents.

\paragraph{Citation Support} Even if a cited document is not relevant or does not address a nugget question, \argue still requires that document to support the sentence that cited it (\autoref{fig:argue}, Judgment D). As  with the nugget attestation check, this is a non-trivial judgment that is obtained via few-shot prompt to the LLM judge. Consistent with \argue's design, our implementation requires that \emph{each} cited document support the \emph{entire} citing sentence.\footnote{A set of citations may support a complex sentence only when considered \emph{jointly}. \argue thus has some bias toward reports with simpler sentences.}

\paragraph{Metrics.} While \argue does not mandate computing specific metrics from the judgments it produces, \citeauthor{mayfield2024evaluation} nonetheless advocate for two metrics, both of which \aargue implements. \emph{Sentence precision} is the proportion of sentences that are attested by \emph{all} of their citations. \emph{Nugget recall} is the proportion of nuggets correctly answered by the report, with a \emph{weighted} variant that weights nuggets by importance (\texttt{okay}=0.5; \texttt{vital}=1.0). Although the nugget attestation check is done per-sentence, we aggregate attested answers across \emph{all} sentences to compute nugget recall.

We additionally compute (un)weighted F1 scores based on these two metrics, which serves as an \emph{overall score}. \aargue also outputs several other fine-grained metrics (e.g.\ \emph{citation support}, the fraction of citations that support their citing sentence).

\paragraph{Rewards and Penalties.} We adopt the following scheme for rewards (green circles) and penalties (red circles):
\begin{itemize}
    \item Each correctly answered nugget (outcomes 3 and 8) counts as an additional (weighted) point in the numerator of the (weighted) nugget recall calculation, where the denominator is simply the (weighted) total number of nuggets for the target topic. Incorrectly answered nuggets do not incur negative points; they merely are not added to the numerator.
    \item Sentences in the sentence precision calculation follow the same logic. Only sentences for which all citations are judged supporting citations (outcomes 2 and 3) are added to the numerator of the sentence precision calculation. The denominator is the total number of sentences deemed to merit a citation---whether they have one or not (outcomes 1, 2, 3, 5).
\end{itemize}

\section{Experiments}
\label{sec:experiments}
\begin{figure}
    \centering
    \includegraphics[width=\linewidth]{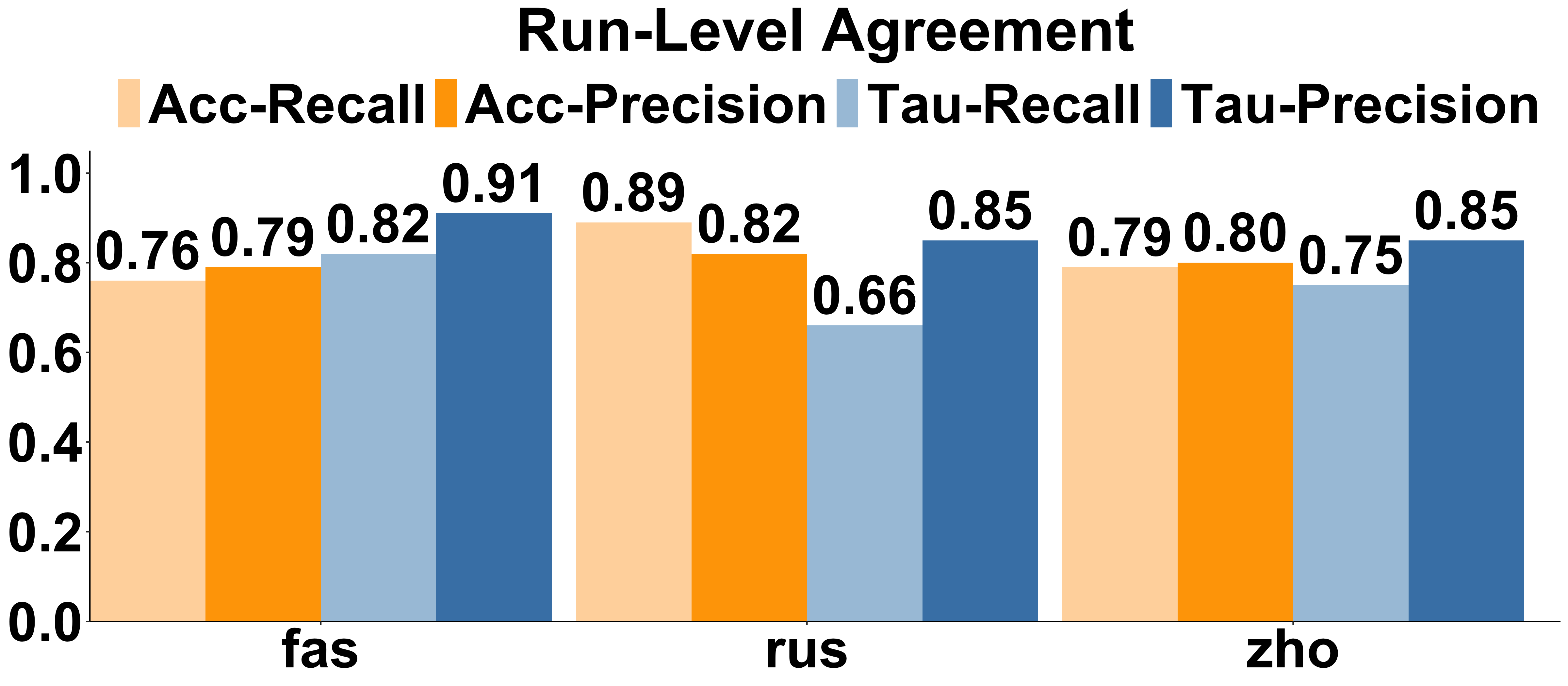}
    \caption{Run-level rank agreement (Kendall's $\tau$ and pairwise accuracy) between \aargue- and human-judged sentence precision (dark shading) and nugget recall (light shading) for the TREC 2024 NeuCLIR report generation pilot.}
    \label{fig:agreement}
    % \vspace{-0.8cm}
\end{figure}

\subsection{TREC NeuCLIR 2024}
\label{sec:experiments::neuclir}

We first evaluate \aargue on the 51 runs from the TREC 2024 NeuCLIR report generation pilot task \cite{lawrie2025overview}, in which English reports are generated based on documents from one of three non-English collections: Chinese, Russian, and Farsi (17 runs each). Human assessors judged sentence precision and nugget recall on reports for the same 21 topics for each run. Topics have 10--20 nuggets, and assessors also identified documents attesting each answer.\footnote{Since a document's relevance in \aargue is determined by whether it attests a nugget answer, we thus also have human-judged document relevance labels.}

Next, we obtain these same metrics from \aargue, using Llama-3.3 70B as the LLM judge \cite{grattafiori2024llama} for all non-trivial judgments in \autoref{fig:argue} (i.e., D, C, G, and H). B judgments use the human relevance assessments. Since all NeuCLIR nuggets are \emph{answerable}, E and F judgments are not generated. We obtain system rankings from (a) assessor-based and (b) \aargue-based macro-average sentence precision and nugget recall across all topics for each language. We compute agreement between these rankings using (i) Kendall's $\tau$ and (ii) the fraction of all pairs of runs for which two run-level Wilcoxon tests---one for the assessor-based ranking and one for the \aargue-based ranking---agree on which of the two runs is superior. \autoref{fig:agreement} presents the results. Broadly, we observe good agreement between the two rankings on both metrics, with particularly strong results on sentence precision. We observe fairly minor cross-lingual variation on sentence precision, but larger variation on nugget recall, which may partly be explained by the variation in nugget answers attested by the documents in each language.

\subsection{TREC RAG 2024}
\label{sec:experiments::rag}

\begin{figure}
    \centering
    \includegraphics[width=0.9\linewidth]{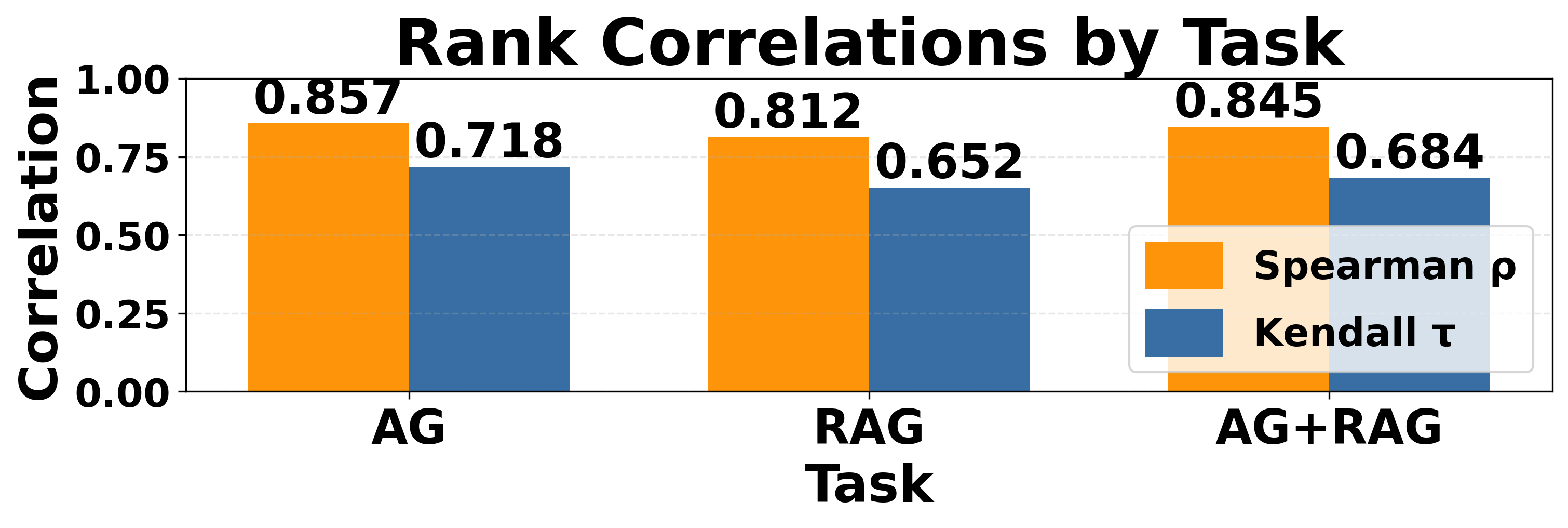}
    \caption{Run-level rank correlations between \aargue's nugget recall and \autonugget's $A_\text{strict}$ metric over submitted to the AG and RAG tasks at TREC RAG 2024.}
    \label{fig:rag-rank-correlations}
    % \vspace{-0.8cm}
\end{figure}

We next compare the nugget recall scores computed by \aargue to similar scores produced by \autonugget \cite{pradeep2025great}, an alternative automatic tool for nugget-based scoring of RAG outputs that was used to assess runs from TREC 2024 RAG track \cite{pradeep2024initial}. We focus on submissions to the ``AG'' and ``RAG'' tasks: the latter requires retrieval of source material, while the former provides a fixed top-$k$ list of source passages from which to generate a response. In total, there are 14 assessor-judged runs for AG and 31 for RAG.

In \autonugget, nuggets are \emph{claims} rather than QA pairs. We use the official nugget sets released by the track organizers, but use Llama-3.3 70B to convert each nugget claim into a QA pair to render them compatible with \aargue.\footnote{This is a simple syntactic transformation that introduces no real semantic changes.}

\begin{figure}
    \centering
    \includegraphics[width=\linewidth]{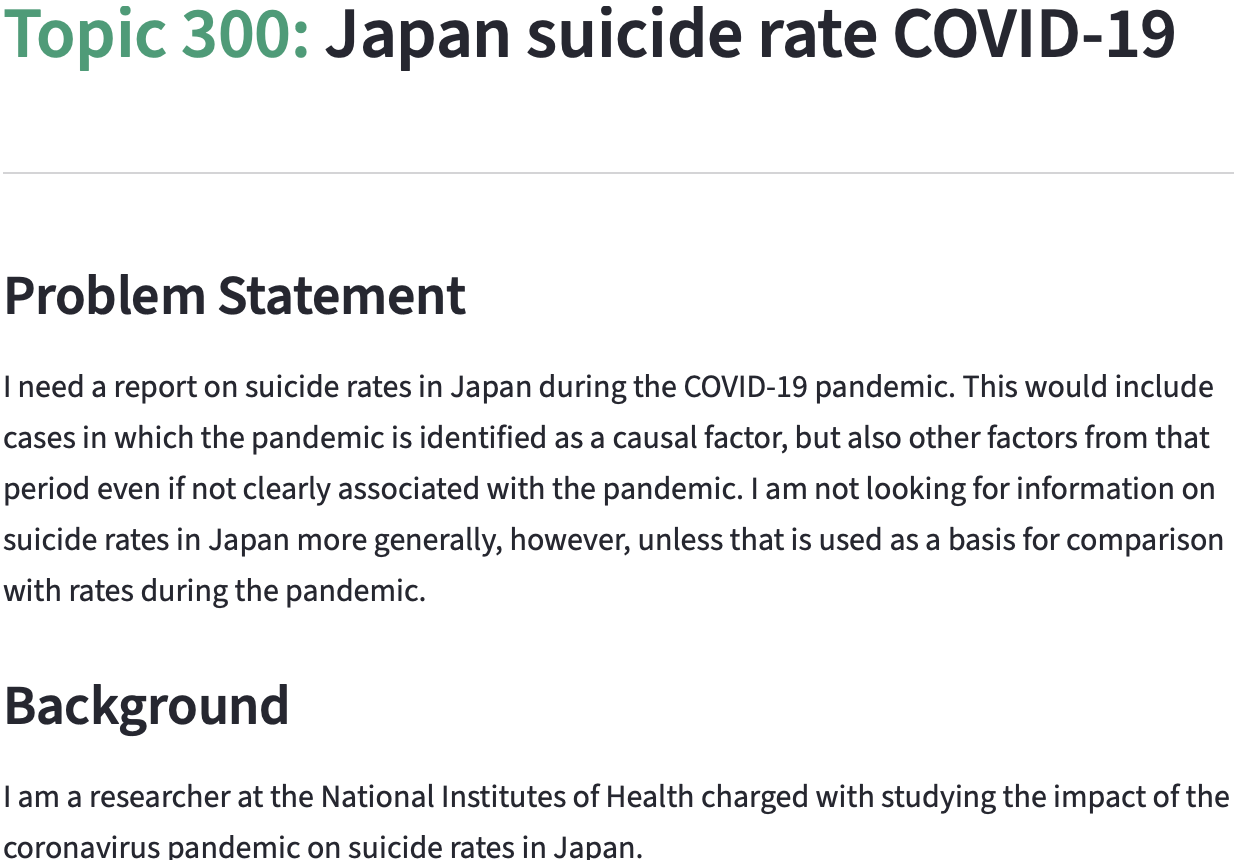}\\~\\
    \includegraphics[width=\linewidth]{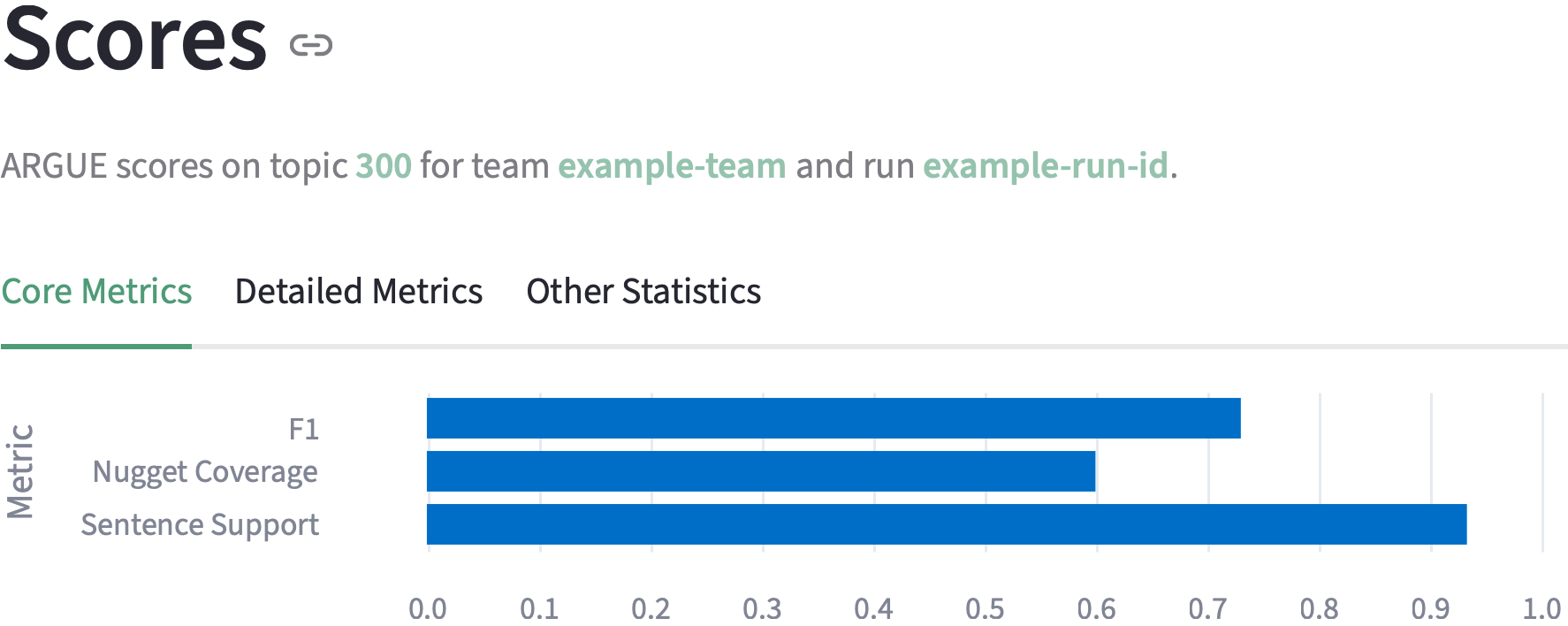}
    \caption{Problem statement, background, and topic-level results in \argueviz for TREC NeuCLIR Topic 300.}
    \label{fig:argueviz-1}
    % \vspace{-0.5cm}
\end{figure}

The official metrics for both the AG and RAG tasks all assess the proportion of nuggets that are attested by the output \cite{pradeep2024initial}, and thus is similar to \argue's nugget recall. The official metrics differ from each other in (1) whether partial support is allowed, (2) which nuggets are included (all nuggets vs.\ only \texttt{vital} nuggets), and (3) whether importance weighting is applied. Here, we focus on their ``$A_\text{strict}$'' metric, which includes \emph{all} nuggets, uses binary support judgments, and does not use importance weighting.\footnote{This choice is largely arbitrary: we obtain very similar results with all official metrics.}

\autoref{fig:rag-rank-correlations} reports run-level correlations for AG, RAG, and all (AG+RAG) runs between rankings induced by \aargue's nugget recall and by \autonugget's $A_\text{strict}$. We find moderate to strong correlations across all three settings, achieving Spearman's $\rho$ of 0.81-0.86 and Kendall's $\tau$ of 0.65-0.72. These findings validate the consistency of \aargue's nugget recall evaluation with similar conceptions of nugget support, as well as its compatibility with nuggets developed in alternative formats.

\section{\argueviz}
\label{sec:argue-viz}
\begin{figure}
    \centering
    \includegraphics[width=\linewidth]{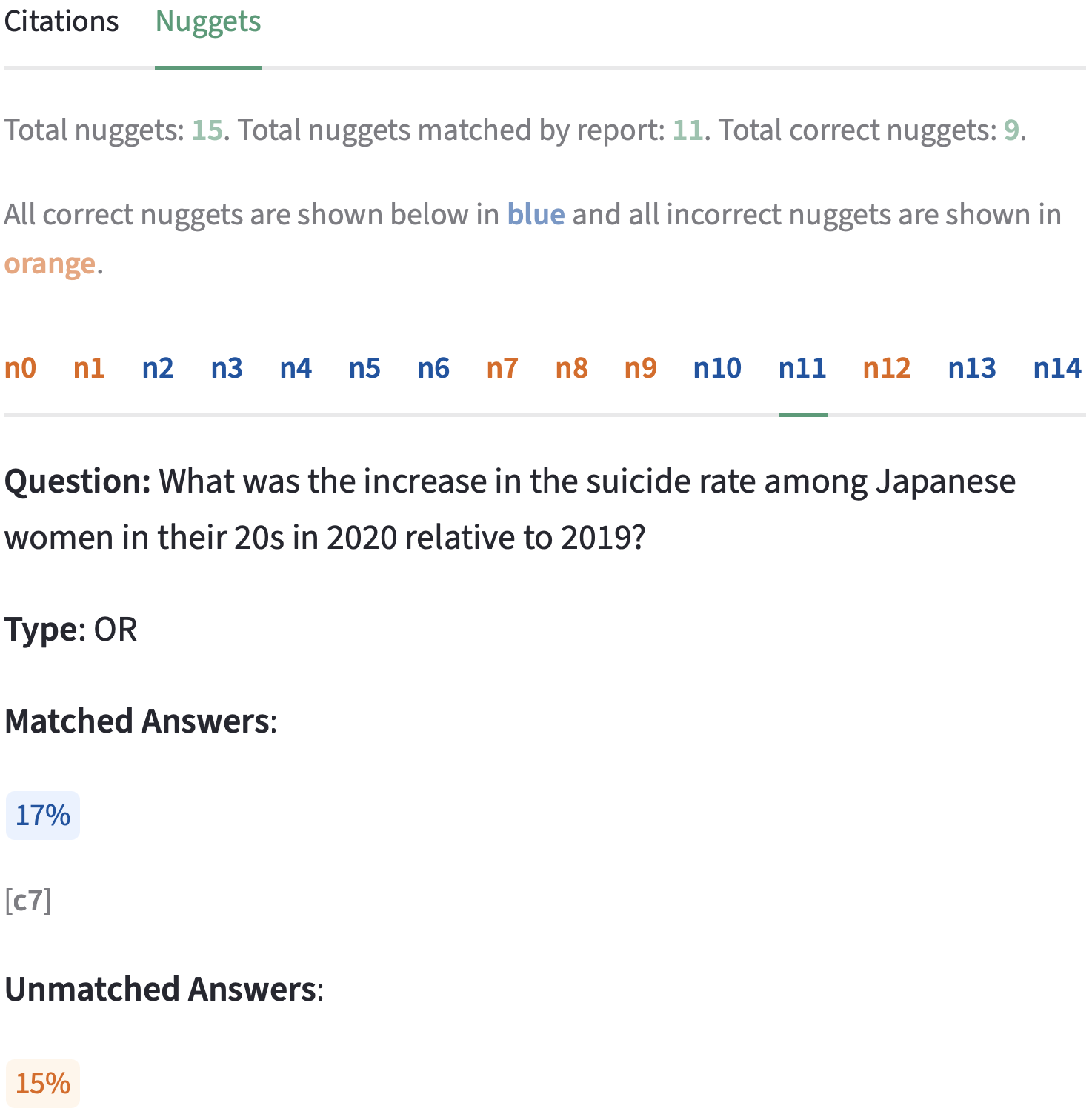}
    \caption{\argueviz report-level view showing (un)attested nugget answers. The ``Citations'' tab (not pictured) shows which report sentences are (un)supported by their citations.}
    \label{fig:argueviz-2}
    % \vspace{-1cm}
\end{figure}

\argueviz is a Streamlit\footnote{\url{https://streamlit.io}} app for visualizing \aargue outputs for a run. Users can toggle between run-level and topic-level results. Run-level results show core metrics (sentence precision, nugget recall, F1) macro-averaged across topics (\autoref{fig:argueviz-1}), while topic-level results display both core and non-core metrics, as well as detailed info about \aargue judgments for that topic.

\aargue judgment details are displayed across two topic-level views: a \emph{report view} (\autoref{fig:argueviz-2}), which shows report-level information about (un)supported sentences and (in)correctly answered nuggets, and a \emph{sentence view} (not pictured) which shows similar information at the sentence level (i.e., which nugget answers are (un)attested by that sentence and which citations support it).

Collectively, these features enable fine-grained error analysis. For example, inspection of correctly and incorrectly answered nuggets on some of our own NeuCLIR runs revealed that our systems consistently struggled to address nuggets requiring multiple answers (i.e., \texttt{AND} nuggets). \argueviz thus can aid both in system development and in diagnosing possible problems with the configuration of \aargue itself.

\section{Conclusion}
\label{sec:conclusion}
This work has introduced \aargue---a configurable LLM-based implementation of the \argue framework for report generation evaluation. Analysis of \aargue on tasks from two tracks at TREC 2024, NeuCLIR and RAG, reveal generally strong run-level correlations between \aargue-based and human assessor-based scores. Further, we introduced \argueviz, a tool for fine-grained inspection of \aargue outputs, enabling detailed error analysis for report generation system development. We release both \aargue and \argueviz to facilitate future work on report generation evaluation.

% \begin{acks}
% \end{acks}

\balance
\bibliographystyle{ACM-Reference-Format}
\bibliography{sample-base}

@inproceedings{mayfield2024evaluation,
  title={On the evaluation of machine-generated reports},
  author={Mayfield, James and Yang, Eugene and Lawrie, Dawn and MacAvaney, Sean and McNamee, Paul and Oard, Douglas W and Soldaini, Luca and Soboroff, Ian and Weller, Orion and Kayi, Efsun and others},
  booktitle={Proceedings of the 47th International ACM SIGIR Conference on Research and Development in Information Retrieval},
  pages={1904--1915},
  year={2024}
}

@inproceedings{es2024ragas,
  title={Ragas: Automated evaluation of retrieval augmented generation},
  author={Es, Shahul and James, Jithin and Anke, Luis Espinosa and Schockaert, Steven},
  booktitle={Proceedings of the 18th Conference of the European Chapter of the Association for Computational Linguistics: System Demonstrations},
  pages={150--158},
  year={2024}
}

@inproceedings{gao2023enabling,
  title={Enabling Large Language Models to Generate Text with Citations},
  author={Gao, Tianyu and Yen, Howard and Yu, Jiatong and Chen, Danqi},
  booktitle={Proceedings of the 2023 Conference on Empirical Methods in Natural Language Processing},
  pages={6465--6488},
  year={2023}
}

@inproceedings{pradeep2025great,
  title={The great nugget recall: Automating fact extraction and rag evaluation with large language models},
  author={Pradeep, Ronak and Thakur, Nandan and Upadhyay, Shivani and Campos, Daniel and Craswell, Nick and Soboroff, Ian and Dang, Hoa Trang and Lin, Jimmy},
  booktitle={Proceedings of the 48th International ACM SIGIR Conference on Research and Development in Information Retrieval},
  pages={180--190},
  year={2025}
}

@inproceedings{saad-falcon-etal-2024-ares,
    title = "{ARES}: An Automated Evaluation Framework for Retrieval-Augmented Generation Systems",
    author = "Saad-Falcon, Jon  and
      Khattab, Omar  and
      Potts, Christopher  and
      Zaharia, Matei",
    editor = "Duh, Kevin  and
      Gomez, Helena  and
      Bethard, Steven",
    booktitle = "Proceedings of the 2024 Conference of the North American Chapter of the Association for Computational Linguistics: Human Language Technologies (Volume 1: Long Papers)",
    month = jun,
    year = "2024",
    address = "Mexico City, Mexico",
    publisher = "Association for Computational Linguistics",
    url = "https://aclanthology.org/2024.naacl-long.20/",
    doi = "10.18653/v1/2024.naacl-long.20",
    pages = "338--354"
}

@inproceedings{voorhees2003overview,
  title={Overview of the TREC 2003 question answering track.},
  author={Voorhees, Ellen M and Dang, Hoa Trang},
  booktitle={Proceedings of the Twelfth Text REtrieval Conference (TREC 2003)},
  volume={2003},
  pages={54--68},
  year={2003}
}

@inproceedings{lin2005automatically,
  title={Automatically evaluating answers to definition questions},
  author={Lin, Jimmy and Demner-Fushman, Dina},
  booktitle={Proceedings of Human Language Technology Conference and Conference on Empirical Methods in Natural Language Processing},
  pages={931--938},
  year={2005}
}

@inproceedings{rajput2011nugget,
  title={A nugget-based test collection construction paradigm},
  author={Rajput, Shahzad and Pavlu, Virgil and Golbus, Peter B and Aslam, Javed A},
  booktitle={Proceedings of the 20th ACM international conference on Information and knowledge management},
  pages={1945--1948},
  year={2011}
}

@incollection{alaofi2024generative,
  title={Generative information retrieval evaluation},
  author={Alaofi, Marwah and Arabzadeh, Negar and Clarke, Charles LA and Sanderson, Mark},
  booktitle={Information access in the era of generative ai},
  pages={135--159},
  year={2024},
  publisher={Springer}
}

@article{grattafiori2024llama,
  title={The llama 3 herd of models},
  author={Grattafiori, Aaron and Dubey, Abhimanyu and Jauhri, Abhinav and Pandey, Abhinav and Kadian, Abhishek and Al-Dahle, Ahmad and Letman, Aiesha and Mathur, Akhil and Schelten, Alan and Vaughan, Alex and others},
  journal={arXiv preprint arXiv:2407.21783},
  year={2024}
}

@article{lawrie2025overview,
  title={Overview of the TREC 2024 NeuCLIR Track},
  author={Lawrie, Dawn and MacAvaney, Sean and Mayfield, James and McNamee, Paul and Oard, Douglas W and Soldaini, Luca and Yang, Eugene},
  journal={arXiv preprint arXiv:2509.14355},
  year={2025}
}

@inproceedings{lajewska2025ginger,
  title={Ginger: Grounded information nugget-based generation of responses},
  author={{\L}ajewska, Weronika and Balog, Krisztian},
  booktitle={Proceedings of the 48th International ACM SIGIR Conference on Research and Development in Information Retrieval},
  pages={2723--2727},
  year={2025}
}

@article{pradeep2024initial,
  title={Initial nugget evaluation results for the trec 2024 rag track with the autonuggetizer framework},
  author={Pradeep, Ronak and Thakur, Nandan and Upadhyay, Shivani and Campos, Daniel and Craswell, Nick and Lin, Jimmy},
  journal={arXiv preprint arXiv:2411.09607},
  year={2024}
}

@article{dietz2026insider,
  title={Insider Knowledge: How Much Can RAG Systems Gain from Evaluation Secrets?},
  author={Dietz, Laura and Li, Bryan and Yang, Eugene and Lawrie, Dawn and Walden, William and Mayfield, James},
  journal={arXiv preprint arXiv:2601.13227},
  year={2026}
}

@article{dietz2026incorporating,
  title={Incorporating Q\&A Nuggets into Retrieval-Augmented Generation},
  author={Dietz, Laura and Li, Bryan and Liu, Gabrielle and Ju, Jia-Huei and Yang, Eugene and Lawrie, Dawn and Walden, William and Mayfield, James},
  journal={arXiv preprint arXiv:2601.13222},
  year={2026}
}

% \appendix

\end{document}